\documentstyle[twoside,fleqn,espcrc2,epsfig]{article}

\newcommand{\AmS}{{\protect\the\textfont2
  A\kern-.1667em\lower.5ex\hbox{M}\kern-.125emS}}

  \newcommand{\beq}{\begin{equation}}
  \newcommand{\eeq}{\end{equation}}

  \newcommand{\beqa}{\begin{eqnarray}}
  \newcommand{\eeqa}{\end{eqnarray}}

  \newcommand{\bit}{\begin{itemize}}
  \newcommand{\eit}{\end{itemize}}
  \newcommand{\nn}{\nonumber }

  \newcommand{\bruch}[2]{{\raisebox{0.4ex}{$#1$}/\raisebox{-0.6ex}{$#2$}}}

  \newcommand{\re}{{\rm Re~}}
  \newcommand{\im}{{\rm Im~}}
  \newcommand{\tr}{{\rm Tr~}}

  \newcommand{\Umat}{\rm 1 \!\!\!\: I}

  \newcommand{\plaq}{\mbox{\raisebox{-2.1mm}
      {\epsfig{file=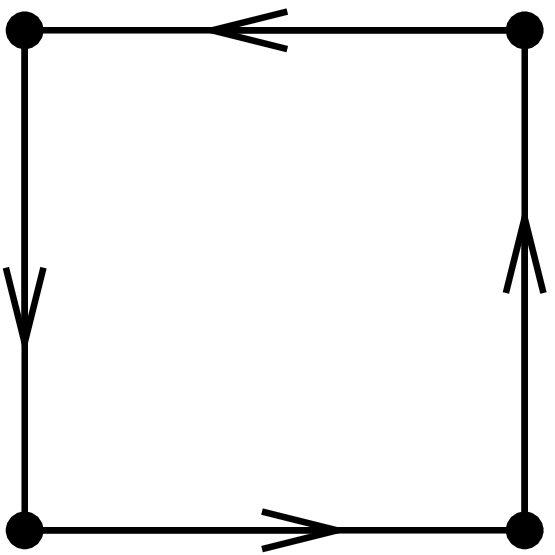,height=8mm
          }}~}}
  \newcommand{\clover}{\mbox{\raisebox{-5mm}
      {\epsfig{bbllx=0,bblly=0,bburx=305,bbury=305,
          file=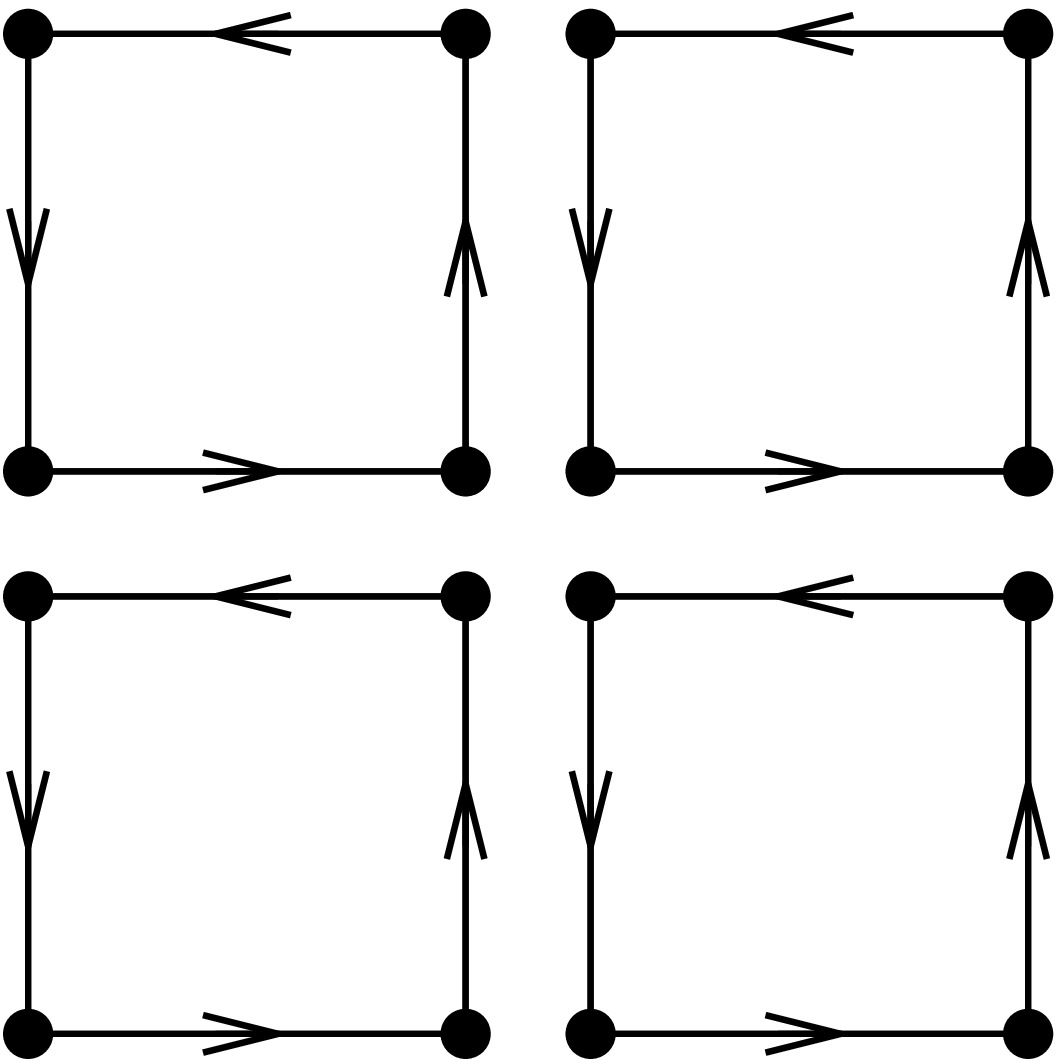,height=13.7mm
          }}~}}
  \newcommand{\link}{\mbox{\raisebox{1mm}
      {\epsfig{bbllx=0,bblly=0,bburx=163,bbury=19,
          file=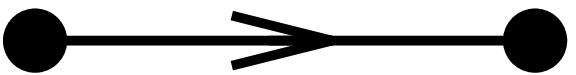,width=12mm
          }}~}}
  \newcommand{\linkdag}{\mbox{\raisebox{1mm}
      {\epsfig{bbllx=0,bblly=0,bburx=163,bbury=19,
          file=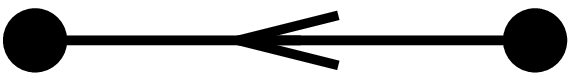,width=12mm
          }}~}}

  \def\NP{{Nucl.\ Phys.\ }}
  
  \def\PL{{Phys.\ Lett.\ }}
  \def\PR{{Phys.\ Rev.\ }}
  
  \def\PRL{{Phys.\ Rev.\ Lett.\ }}

\title{Finite Temperature Meson Masses with Improved Quenched Wilson Fermions}

\author{P. Schmidt\address{
Fakult\"at f\"ur Physik, Universit\"at Bielefeld, 33501 Bielefeld, Germany}
 with $\mbox{E. Laermann}^{\mbox{\scriptsize a}}$  
       }

\begin{document}

\begin{abstract}
We present recent results for meson screening masses with
quenched Wilson fermions above and below the confinement-
deconfinement phase transition. The action used in the
simulation is the Sheikholeslami Wohlert action. The quark
masses are chosen to be in a range going from light quarks
up to the mass of the charm quark. These results are compared 
with zero temperature Wilson and high temperature
staggered data.
\end{abstract}

\maketitle

\section{Introduction}

Mesons in a wide range of the quark mass have been thoroughly investigated 
throughout the last years in quenched studies with various kind of actions 
by many groups and collaborations. These investigations have mostly 
concentrated on zero temperature physics to extract the mass spectrum. 
Some studies have beeen performed on high temperature meson screening 
masses a few years ago by some groups\cite{Gocksch,Born,Gupta,Boyd} 
using quenched staggered quark fields.

This study has been made to examine 
the behaviour of the pseudoscalar and the vectormeson screening mass in the
quenched Wilson formulation slightly
below and above the high temperature phase transition. 
Furthermore we computed
the free quark propagator analytically to see wether the pion and the rho
are approaching the 
high temperature limit of $m_{\pi,\rho} = 2\sqrt{ (\pi T)^{2}  + m_{q}^{2}}$.

\section{Simulation}

\subsection{Action}

We have used the standard Wilson action for the pure gauge part 
\beqa 
S_{Gluon} & = & \frac{6}{g^2} \sum_{x,\mu > \nu} 
\left( 1 - \frac{1}{N} \re\tr\plaq_{\!\mu\nu}(x) \right) \nn
\eeqa
and the Sheikholeslami Wohlert action\cite{Sheikholeslami} for the fermion part
setting the clover coefficient  to $C_{sw} \! = \! 1$.\\
\begin{eqnarray*}
\lefteqn{S_{Clover} \;\;  =  \;\; \frac{1}{2 \kappa} \sum_{x,y} \bar{\Psi}(x) \Bigg\{ 
} \\
&& \hspace*{-5mm} \Bigg( \Umat - \frac{\kappa\:\!C_{sw}}{2}\sum_{\mu,\nu}
           \im\clover_{\!\!\mu\nu}\!(x)\,\sigma_{\mu\nu}(x) \Bigg)\delta_{x,y}
\nn \\
&& \hspace{-3mm} - \,\,\kappa\,\, \sum_{\mu}\Bigg[
          (\Umat - \gamma_\mu )\,\delta_{x+\hat{\mu},y}\;\link_{\!\!\mu}(x) +
\nn \\
&& \hspace{12.5mm}  (\Umat + \gamma_\mu )\,\delta_{x-\hat{\mu},y}\;\linkdag_{\!\!\mu}(y)
          \Bigg] \Bigg\} \Psi(y)  .\nn 
\end{eqnarray*}

\subsection{Observables}

We have used extended sinks for the meson correlation function 
the way Lacock et al.\cite{Lacock} proposed in 1995.
Hence we have been able to fit meson propagators over a much wider range 
than in the local case to get more reliable results. 
The most suitable distance between quark and antiquark 
has been between 3 and 8 lattice units 
depending on the quark mass and the lattice spacing. 

We have computed the meson propagators by combining two 
quark propagators generated with equal or different $\kappa$-values
to obtain a large set of masses  
$ m_{q} = \frac{1}{2} \, (m_{q_{1}} + m_{q_{2}})$ with $m_{q_{i}} = 
\ln (1 + \frac{1}{2}( \frac{1}{\kappa_{i}} - \frac{1}{\kappa_{c}}))$.

\subsection{Technical details}

The simulations were performed at temperatures 
somewhat below the transition temperature 
phase transition at about $T = 0.9 \; T_{C}$ on
$16^{3} \! \times \! 8$, $24^{3} \! \times \! 8$ and 
$32^{3} \! \times \! 8$ lattices at $\beta = 6.0$ 
and above the transition at about $T = 1.2 \; T_{C}$
on a $24^{3} \! \times \! 8$ lattice at $\beta = 6.2$. The lattice spacings are
$a = 2.05(6) \mbox{GeV}^{-1}$ and $a = 2.7(1) \mbox{GeV}^{-1}$ 
computed from zero temperature data
obtained from the literature\cite{Allton1,Allton2,Allton3,Baxter}.

We also performed simulations on zero temperature lattices\cite{schmidt}
at $\beta \! = \! 6.0 \; (16^{3} \! \times \! 32)$ and 
$\beta \! = \! 6.2 \; (24^{3} \! \times \! 48)$ 
to make sure that we are consistent with other groups.

The number of configurations lies between 20 and 100 being 40 or 100
in most cases.

\section{Results}

\subsection{\bf $\mbox{T} = \mbox{0.9} \; \mbox{T}_{C}$}

We first compare the finite temperature results near $\kappa_{c}$
at $T=0.9 \; T_{C}$ with the zero temperature data (Fig. 1). 
The dotted and dashed lines are linear fits.
For the pseudoscalar meson 
the line of the high temperature screening mass 
comes to lie nearly on top of the zero temperature line. The small deviation 
is possibly due to finite size effects. 
Both sets of data extrapolate to nearly the same point of vanishing meson mass.

For the vectormeson channel we observe a larger deviation between
the high temperature and the $T \! = \! 0$ data,
even if we shift $\kappa_{\mbox{\tiny $T \! = \! 0$}}$ and 
$\kappa_{\mbox{\tiny $T \! = \! 0.9T_{c}$}}$ indicated 
by the vertical lines on top of each other. 
This trend is perhaps a bit more pronounced than in results obtained
with staggered fermions \cite{Boyd}. 
\vspace*{3.5mm}

\subsection{$ \mbox{T} = \mbox{1.2} \; \mbox{T}_{C}$}\label{T=1.2Tc}

We then compare the finite temperature results 
at $T \! = \! 1.2 \; T_{C}$ with the zero temperature data over a wide range 
of the quark mass going from light masses near the chiral limit up to masses
near $m_{\mbox{\scriptsize charm}}$ (Fig. 2).
The zero temperature curves are fits to $ a + bx + cx^{2} $ while the 
high temperature curves are polynomials to guide the eye.

In contrast to the results at $T \! = \! 0.9 \; T_{C}$ the screening masses 
of the 
pseudoscalar and the vectormeson aquire a large shift upwards compared to 
the zero temperature data. The pion remains heavy even in the chiral limit
by restoration of the chiral symmetry above the phase transition temperature.
\vspace*{-6mm}
\begin{center}
{\epsfig{file=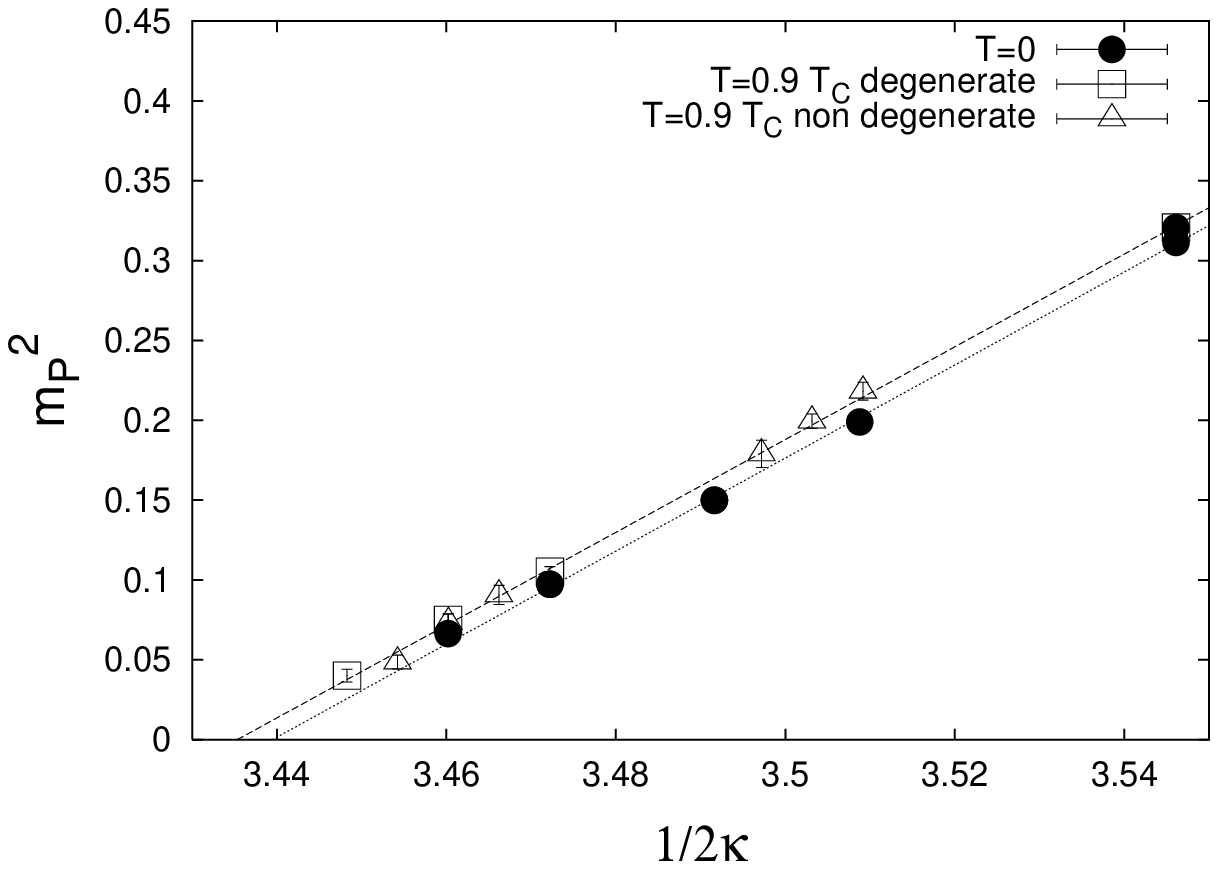,
height=5.4cm}}\\
{\epsfig{file=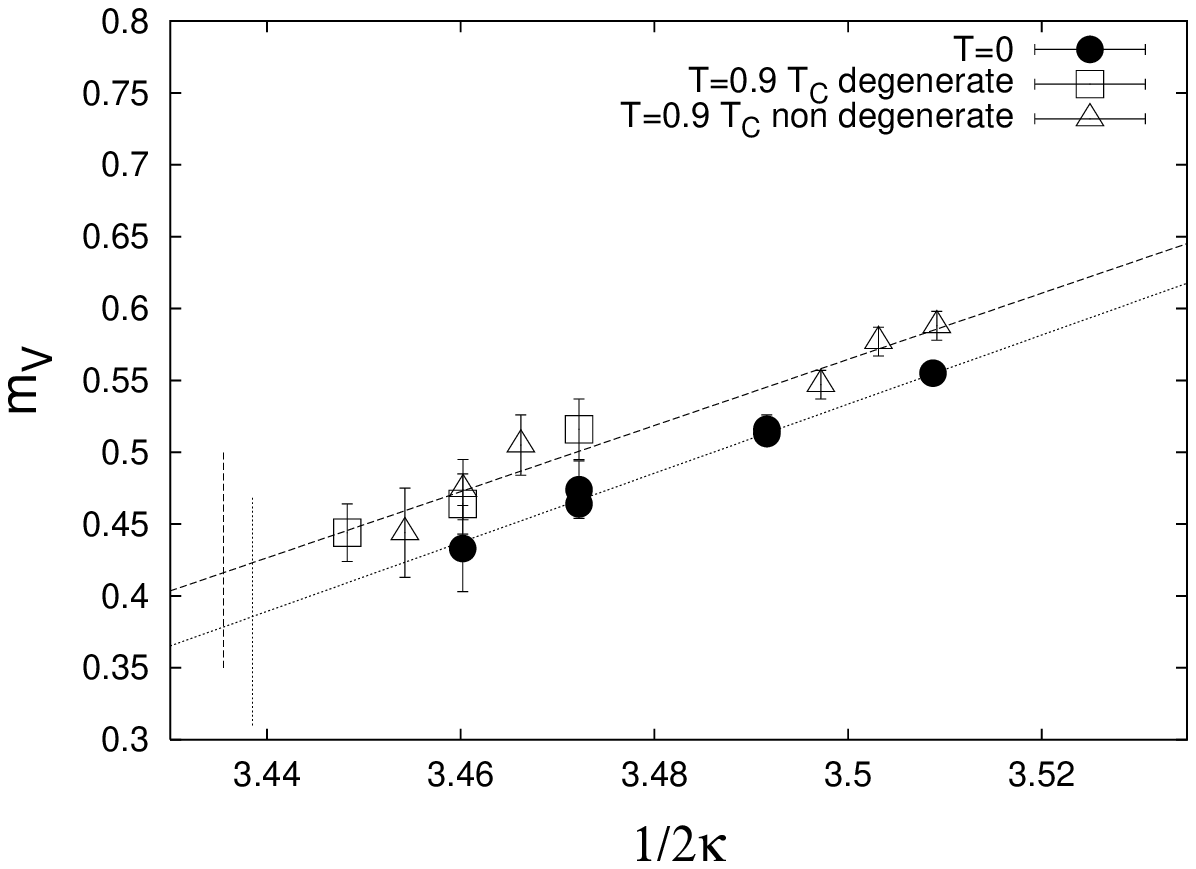,
height=5.4cm}}
\end{center}
\vspace*{-2mm}
{\bf Figure 1:} 
{\small The mass of the pseudoscalar squared and the mass of the vectormeson
as a function of  $\bruch{1}{2\kappa}$ near $\kappa_{c}$ at $\beta = 6.0$ for
$T = 0.9 \; T_{C}$ (open symbols) and zero temperature (filled symbols).}

\vspace*{5mm}

The rhomass $(m_{\rho} \! \approx \! 0.77 a^{-1})$ is still heavier 
than the pionmass $(m_{\pi} \! \approx \! 0.64 a^{-1})$, but they both
seem to approach the high temperature limit of two free propagating 
quarks (see \ref{The free field}). 

This result is in agreement with staggered data on quenched meson 
screening masses \cite{Born}.


\subsection{The free field}\label{The free field}
On a finite lattice we have analytically 
computed the free fermion field which is
believed to be the high temperature limit $(\beta \! \rightarrow \! \infty)$.
The summation over the internal momenta has been done numerically.
The results are shown in figure 3.

The results for the $24^{3} \! \times \! 8$ lattice are indicated by 
the dashed dotted line and the infinite volume limit by the
fat line. 
On the $24^{3} \! \times \! 8$ lattice the mass of the pion and the rho for 
$m_{q} \! \rightarrow \! 0$ are
$m_{\pi,\rho} \! \approx \!  0.85a^{-1}$ while the 
continuum value is $m_{\pi,\rho} \! = \! 2 \pi / 8 \! = 0.785a^{-1}\!$.
Thus on finite lattices the high temperature limit is approached from above. 
The results of the numerical simulation are quite close to this number
at $T \! = \! 1.2\;T_{C}$ already.

\begin{center}
{\epsfig{file=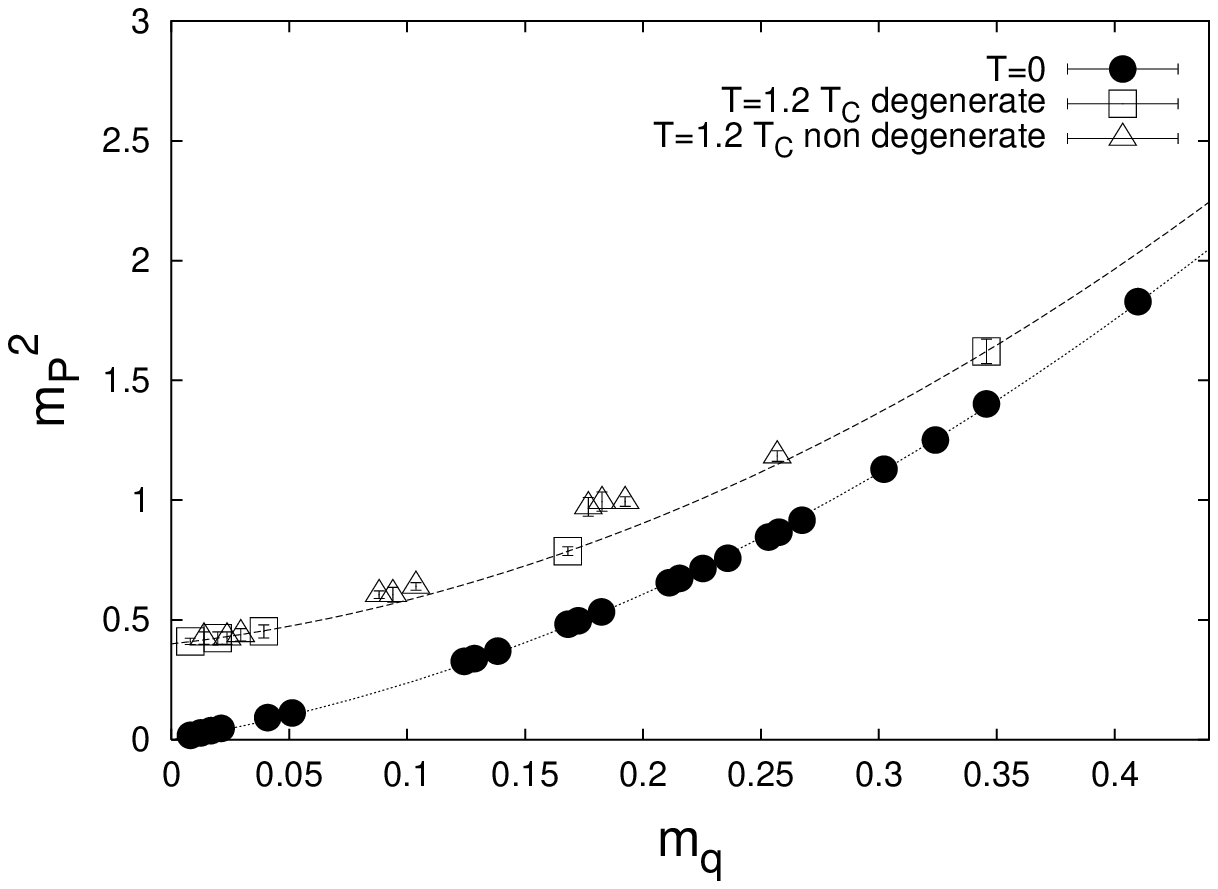,
height=5.4cm}}\\
{\epsfig{file=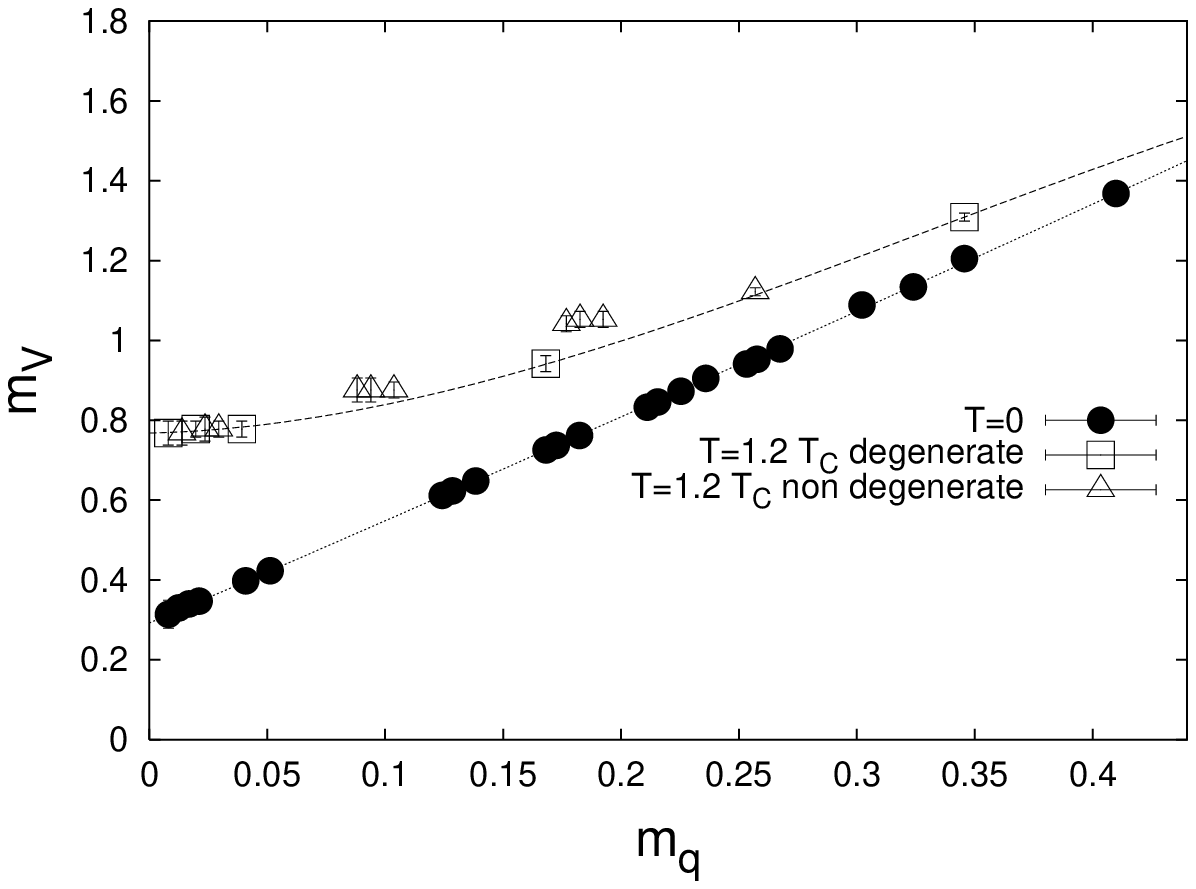, 
height=5.4cm}}
\end{center}
\vspace*{-2mm}
{\bf Figure 2:}
{\small The mass of the pseudoscalar squared and the mass of the vectormeson
as a function of  $m_q = \bruch{1}{2}(m_{q_{1}} + m_{q_{2}})$ at $\beta = 6.2$ for
$T = 1.2 \; T_{C}$ (open symbols) and zero temperature (filled symbols).
}\\

In addition to that we have looked at the different behaviour of the 
degenerate and non degenerate quark mass meson propagators. 
The non degenerate meson masses in the 
infinite volume limit (dotted line) lie
significantly above the degenerate masses, which confirmes the results 
of the simulations (squares and triangles). The dashed curve is a 
polynomial to guide the eye where the degenerate meson masses are 
expected to be.
The behaviour is understood as the influence of the two quarks being
separately propagating 
\{$\; 2\sqrt{( \pi T)^{2}  + m_{q}^{2}} \leq \sqrt{( \pi T)^{2}  + 
m_{q_{1}}^{2}} + \sqrt{( \pi T)^{2}  + m_{q_{2}}^{2}} \; , \;  
(m_{q} = \bruch{1}{2} (m_{q_{1}} +m_{q_{2}}))\; $\}. 
\begin{center}
{\epsfig{file=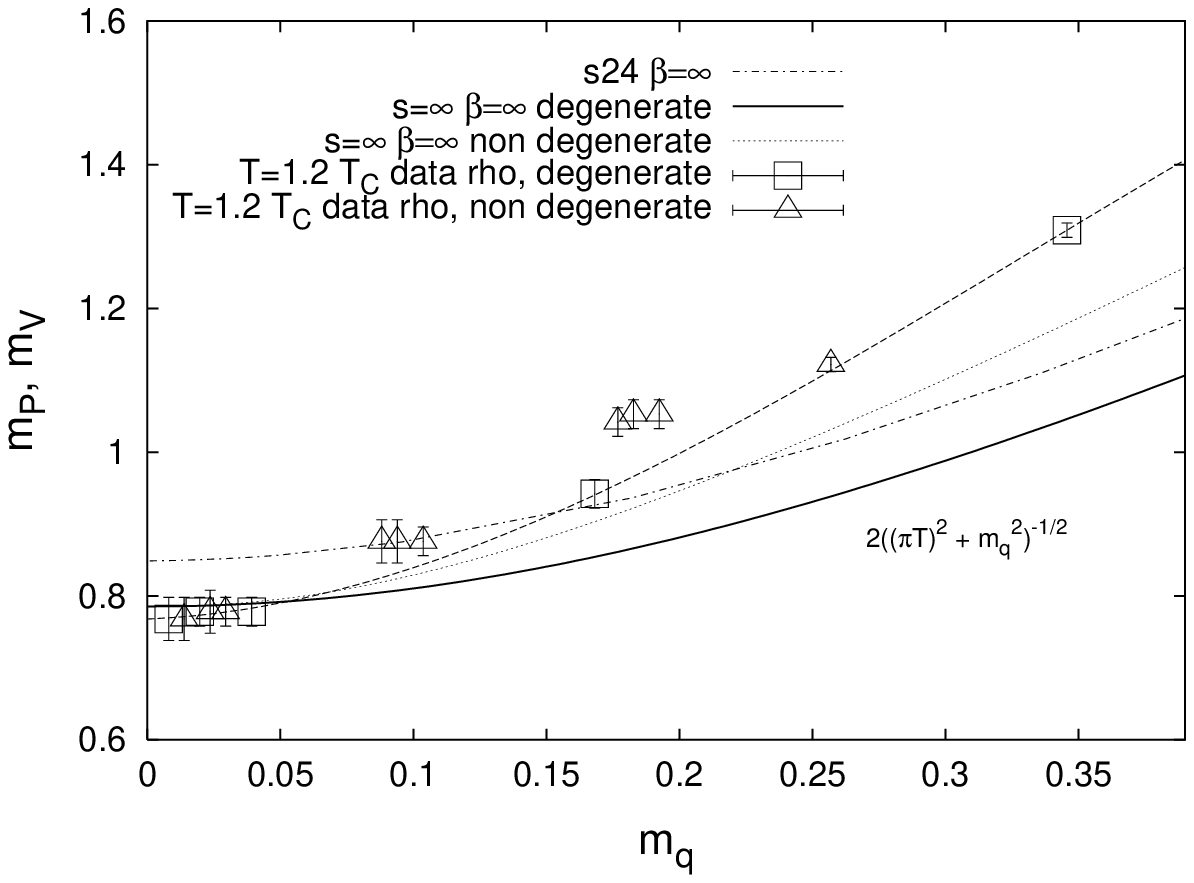, 
height=5.4cm}}
\end{center}
\vspace*{-2mm}
{\bf Figure 3:}
{\small The masses of the vectormeson 
for $T=1.2 \; T_{C}$ against the bare quark mass including
the results for the free quark field $(\beta \! = \! \infty)$ on 
a $24^{3} \! \times \! 8$ lattice and the infinite volume limit with
equal and different constiuent masses. 
}\\



\begin{thebibliography}{9}

\bibitem{Gocksch}
A. ~Gocksch et al., \\ 
\PL B205 (88) 334

\bibitem{Born}
K. D. ~Born et al., \\ 
\PRL 67 (91) 302

\bibitem{Gupta}
S. ~Gupta  \\ 
\PL B288 (91) 302
 
\bibitem{Boyd}
G. ~Boyd et al., \PL B349 (95) 170

\bibitem{Sheikholeslami} 
B. ~Sheikholeslami and R. ~Wohlert, \\
\NP B259 (1985) 572

\bibitem{Lacock}
UKQCD Collaboration (P. ~Lacock et al.), \\
\PR D51 (1995) 6403 

\bibitem{Allton1}
UKQCD Collaboration (C.R. ~Allton et al.), \\
\NP B407 (1993) 331

\bibitem{Allton2}
APE Collaboration (C.R ~Allton et al.), \\
\NP B413 (1994) 461

\bibitem{Allton3}
UKQCD Collaboration (C.R. ~Allton et al.), \\
\PR D49 (1994) 474 

\bibitem{Baxter}
UKQCD Collaboration (R.M. Baxter et al.), \\
\PR D49 (1994) 1594


\bibitem{schmidt}
P. ~Schmidt, ~Diploma Thesis

\end{thebibliography}
\end{document}